\documentclass[mathleft]{an}
\usepackage{graphicx}
\usepackage{times}
\usepackage{amsmath}
\usepackage{amssymb}
\begin{document}

\Pagespan{789}{}
\Yearpublication{2006}%
\Yearsubmission{2005}%
\Month{11}%
\Volume{999}%
\Issue{88}%

\title{Termination of star formation by BH feedback in equal- and unequal-mass
  mergers of disk and elliptical galaxies}

\author{Peter H. Johansson\inst{1}\fnmsep\thanks{Corresponding author:
  {pjohan@usm.lmu.de}}, Thorsten Naab\inst{1}, Andreas Burkert\inst{1}}
\titlerunning{Termination of star formation by BH feedback in galaxy mergers}
\authorrunning{Johansson et al.}
\institute{
University Observatory Munich, Scheinerstr.\ 1, D-81679 Munich, Germany}

\received{22 August 2008}
\accepted{}
\publonline{later}


\abstract{%
We present binary galaxy merger simulations of gas-rich disks
(Sp-Sp), of early-type galaxies and disks (E-Sp, mixed mergers), and mergers of early-type galaxies 
(E-E, dry mergers) with varying mass ratios and different progenitor morphologies. The simulations
include radiative cooling, star formation and black hole (BH) accretion and
the associated feedback processes. We find for Sp-Sp mergers, that 
the peak star formation rate and BH accretion rate decrease
and the growth timescales of the central black holes and newly formed stars
increase with higher progenitor mass ratios. The termination of star formation
by BH feedback in disk mergers is significantly less important
for higher progenitor mass ratios (e.g. 3:1 and higher). In addition, the
inclusion of BH feedback suppresses efficiently star formation in dry E-E mergers and mixed E-Sp mergers.}

\maketitle

\section{Introduction}

Recent large-scale statistical galaxy surveys (e.g. the Sloan Digital Sky Survey (SDSS) and
the two-degree Field (2df) survey) have revealed a robust bimodality in the observed galaxy population.
Galaxies above a critical stellar mass of $M_{\rm crit}\simeq 3\times 10^{10} M_{\odot}$ are 
typically non-star forming red spheroidal systems with old stellar populations that predominately live 
in dense environments, whereas galaxies below this critical mass are typically blue, star-forming 
disk galaxies that lie in the field (e.g. Bell et al. 2003; Kauffmann et
al. 2003; Baldry et al. 2004; Balogh et al. 2004).

The observed bimodality can also be seen as a manifestation of cosmic
downsizing, in which the most massive galaxies formed a significant proportion
of their stars at high redshifts above $z \gtrsim 2$.
The star formation is efficiently quenched in these systems by $z=1$ as 
manifested in the observed population of extremely red non-starforming massive ellipticals 
(EROs, e.g. McCarthy 2004; V\"ais\"anen \& Johansson 2004).
The lower mass systems, on the contrary, have 
typically been forming stars throughout the whole cosmic epoch 
(e.g. Glazebrook et al. 2004; Juneau et al. 2005; Cimatti, Daddi, Renzini 2006).

The quenching mechanism responsible for the observed termination of star formation in massive galaxies at 
redshifts below $z \lesssim 2-3$ needs to be both energetic enough to trigger the quenching and long-lasting 
enough to maintain the quenching over a Hubble time. There exists several theoretical explanations for the quenching, 
including the feedback from AGNs (e.g. Bower et al. 2006; Croton et al. 2006),
gaseous major mergers triggering star burst and/or quasar activity 
(e.g Naab, Jesseit, Burkert, 2006), quenching by shockheated gas above a critical halo mass 
(Dekel \& Birnboim 2006; Birnboim, Dekel, Neistein 2007) and gravitational quenching by clumpy accretion 
(Naab et al. 2007; Dekel \& Birnboim 2008).

In this paper we study the termination of star formation by black hole (BH) feedback in binary galaxy mergers.
The study by Springel, Di Matteo, Hernquist (2005a) showed that BH feedback
efficiently terminates star formation in equal-mass disk 
mergers. Here we confirm this result and expand the analysis to include unequal-mass disk mergers, 
mixed E-Sp mergers and dry E-E mergers. This paper is structured as follows. In \S \ref{Sims} we briefly describe our simulation code and galaxy merger setup.
We discuss the effect of varying the initial gas fractions, orbits, merger mass ratios and galaxy progenitors 
for terminating star formation in \S \ref{Results}. Finally, we summarize and discuss our findings \S \ref{Conc}.

\section{Simulations}
\label{Sims}

\begin{figure}
\centering 
\includegraphics[width=8.2cm]{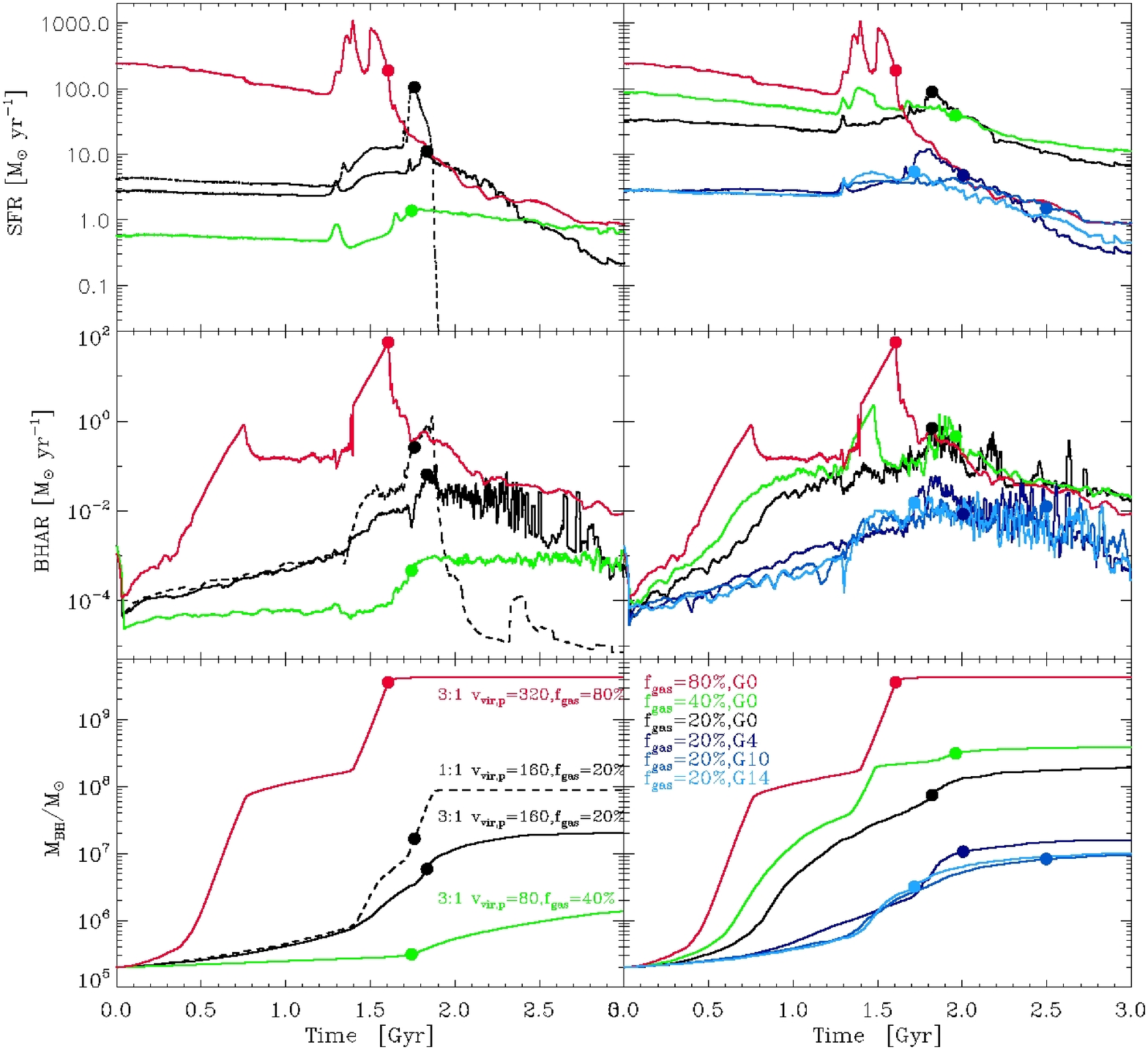}
\caption{The total star formation rate (top), the total black hole accretion rate (middle)
and the evolution of the total black hole mass (bottom) as a function of time for
three 3:1 (solid lines) and one 1:1 (dashed line) merger with initial gas mass 
fractions of 20\% (black), 40\% (green) and 80\% (red) \textit{(left panel)} and for three 
3:1 mergers with varying initial gas mass fractions (black, green, red
lines) and three 3:1 mergers with varying orbital and initial geometries for a
fixed gas fraction (blue lines) \textit{(right panel)}. The filled circles
indicate the time of merging of the BHs.}
\label{31_sfr}
\end{figure}

We perform the simulations using the parallel TreeSPH-code 
GADGET-2 (Springel 2005) including star formation and the associated 
supernova feedback, where the multiphase interstellar medium 
(Efstathiou 2000; Johansson \& Efstathiou 2006)
is modeled using the sub-resolution
mo-del developed by Springel \& Hernquist (2003). We implemented a black hole feedback 
description similar to the Sp-ringel et al. (2005b)
model into our version of the code.
In this model the BH accretion rate is parameterized using a Bondi-Hoyle-Lyttleton description, 
with the maximum accretion rate limited to the Eddington rate. 
Finally, some fraction of the accreted rest mass energy is available as thermal feedback energy
that couples to the surrounding gas. Further details about the code implementation and 
parameter choices can be found in Johansson, Naab, Burkert (2008), J08 hereafter. 

The disk galaxies are setup as described in J08, with Hernquist dark matter profiles populated 
by exponential dis-ks with initial gas fractions ranging from 20\% to 80\%. All models contain initially
a bulge with a third of the total disk mass. We also model mergers of early-type galaxies, where the early-type 
galaxies are setup initially using merger remnants of binary disk mergers
(Naab \& Burkert 2003). We set the 
primary galaxy models up with 20,000 gas and stellar disk particles, 10,000 bulge particles and 
30,000 dark matter particles and scale the other models correspondingly.
The gravitational softening lengths of gas, newly formed stars and the
black hole particles are set to $\epsilon=0.1 h^{-1} \rm{kpc}$ and the softening
lengths of the disk, bulge and more massive dark matter particles are scaled with the
square root of the particle masses resulting in  $\epsilon=0.2 h^{-1} \rm{kpc}$
for the bulge and disk particles and $\epsilon=0.8 h^{-1} \rm{kpc}$ for the
dark matter particles, respectively. 

The galaxies are initially placed on parabolic orbits wh-ere the initial separation
of the progenitors is set to the mean of the two galaxy virial radii and the pericentric distance to the 
mean of the two disk scale radii. All simulations presented in this paper were evolved for a total of $t=3 \ \rm
Gyr$ using the local Altix 3700 Bx2 machine hosted at the University Observatory in Munich.

\section{Results}
\label{Results}

In the left panel of Fig. \ref{31_sfr} we plot the total star formation rates, BH accretion rates and
the BH mass growth for three 3:1 mergers (solid lines) with  20\% (black), 40\% (green), 80\%
(red) initial gas mass fractions together with one 20\% gas fraction 1:1 merger (dashed lines).
The star formation is very efficiently terminated by the BH feedback in 1:1 mergers, compared
to a generally much shallower decline in star formation for 3:1 mergers. In addition, the final 
BH mass are lower in 3:1 mergers typically by a factor of 2-5, but with a relatively large scatter 
depending on the progenitor masses and initial gas fractions. 

In the right panel of Fig. \ref{31_sfr} we show the results for three 3:1 mergers
with varying initial gas mass fractions for
a fixed orbit and initial disk geometry (black, green, red lines) and for three 3:1 
mergers with varying initial orbits and orientations for a fixed gas mass fraction (blue lines). 
The initial gas fraction has a large effect on the height of the star
formation and BH accretion peaks, with larger initial gas fractions producing
higher values, as expected. This results also in relatively large differences 
in the final BH masses, with the  $f_{\rm gas}=0.8$ simulations producing
final BH masses that are larger by an order of magnitude compared to the 
$f_{\rm gas}=0.2$ runs. The variation of the orbit and initial geometry
for a fixed gas mass fraction produces much smaller differences. The peaks of the 
star formation rates and BH accretion rates only vary within a factor of two
with changing orbits and initial disk geometries. 

\begin{figure}
\centering 
\includegraphics[width=6.5cm]{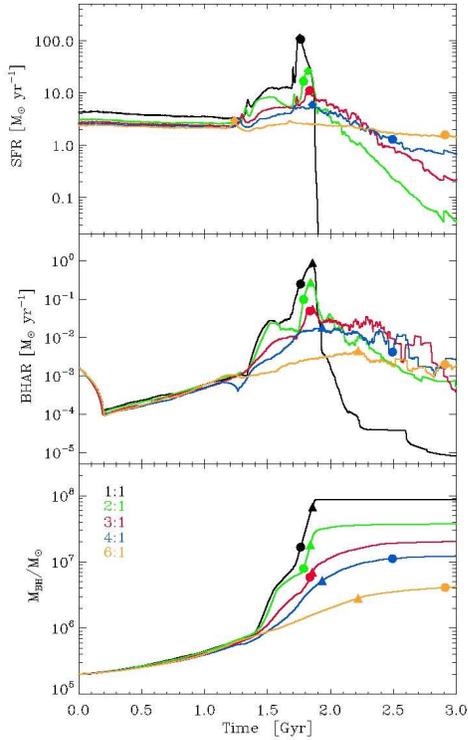}
\caption{The evolution of the star formation rates, black hole accretion rates and BH mass
as a function of time for
co-planar prograde 1:1 (black), 2:1 (green), 3:1 (red), 4:1 (blue) and 6:1
(orange) mergers. The filled circles indicate the time of the BH merger, the
filled diamonds in the top panel and the filled triangles in the bottom two panels
show the location of the maximum star formation and BH accretion peaks, respectively.}
\label{SFR_BHacc_uneq}
\end{figure}

In Fig.\ref{SFR_BHacc_uneq} we study the star formation and BH accretion histories for
unequal-mass mergers with varying mass ratios. To this end we ran five mergers with
mass ratios of 1:1, 2:1, 3:1, 4:1 and 6:1 on co-planar prograde orbits with initial gas mass
fractions of 20\%. As can be seen in Fig. \ref{SFR_BHacc_uneq} increasing the mass ratio of 
the merger systematically lowers the
peak star formation rate and increases the duration of star formation activity
after the merger. For the highest mass ratio merger the star formation rate is virtually
constant throughout the simulation with only a mild peak during the first
passage (see also di Matteo et al. 2007).
The final BH masses are systematically lower for increasing mass
ra-tio of the merger. Furthermore, the slope of the $M_{\rm{BH}}$ growth as a
function of time becomes shallower with increasing merger mass ratio. 
There is also a systematic delay in the time of the BH merger with increasing mass
ratio, as indicated by the filled circles in Fig. \ref{SFR_BHacc_uneq}. For
the lower mass ratio mergers the peak of the BH activity (the filled triangles
in Fig. \ref{SFR_BHacc_uneq}) typically occurs shortly after the merging of
the BHs, whereas for the higher mass ratio mergers the peak of the BH activity
is not directly related to the merging time of the BHs.

\begin{figure}
\centering 
\includegraphics[width=6.5cm]{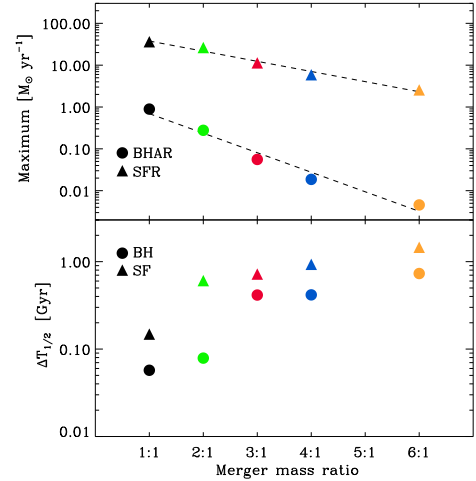}
\caption{The maximum BH accretion and star formation rates and timescales as a
  function of merger mass ratio. The bottom panel shows the half mass growth time of the BHs (circles)
and the stellar mass (triangles) centered on the maximum BH accretion
and star formation rates, respectively. The top panel shows the corresponding
peak BH accretion (circles) and star formation rates (triangles) during the
corresponding half mass growth time  $\Delta{T_{1/2}}$.}
\label{SFR_BH_acctime_uneq}
\end{figure}

In Fig. \ref{SFR_BH_acctime_uneq} we study the duration of the BH accretion
and star formation activity as a function of merger mass ratio. We define a
half-mass growth time $\Delta{T_{1/2}}$ during which half of the final BH mass
and half of the total stellar mass is formed respectively. In both cases
$\Delta{T_{1/2}}$ is centered on the peak of the corresponding activity, the
BH accretion on the maximum of the BH accretion rate (the triangles in
Fig. \ref{SFR_BHacc_uneq}) and the star formation rate on the peak of the SFR
marked with diamonds in Fig. \ref{SFR_BHacc_uneq}. 
The resulting half-mass timescales are shown in the bottom panel of Fig. \ref{SFR_BH_acctime_uneq}.  
The $\Delta{T_{1/2}}_{\rm BH}$ is strongly correlated with the mass
ratio of the merger. For 1:1 and 2:1 mergers the growth of the $M_{\rm{BH}}$
is very concentrated in time, with half of the final BH mass growth occurring in less
than 100 Myr. For the 3:1 and 4:1 mergers $\Delta{T_{1/2}}_{\rm BH}\sim 0.5 \ \rm
Gyr$, with the 6:1 merger resulting in $\Delta{T_{1/2}}_{\rm BH}\sim 1 \ \rm
Gyr$. 

The corresponding stellar half-mass timescales $\Delta{T_{\frac{1}{2}}}_{\rm star}$ (triangles in
the bottom panel Fig. \ref{SFR_BH_acctime_uneq}) also show a clear correlation with the mass ratio of the
merger. In the 1:1 merger half of the final stellar mass is formed in a short
major burst lasting about $\Delta{T_{\frac{1}{2}}}_{\rm star}\sim 0.15 \ \rm Gyr$. This
value is comparable to the star formation timescale derived 
by Cox et al. (2008), 
who calculated a full width at half maximum of
$\rm{FWHM}\sim 0.1 \ \rm{Gyr}$ for the star formation peak of a typical 1:1 merger. 
For higher mass-ratio mergers the resulting star formation
timescales are $\Delta{T_{\frac{1}{2}}}_{\rm star}\sim 0.7-1.0 \ \rm{Gyr}$, with the highest
mass-ratio merger having the longest timescale of $\Delta{T_{\frac{1}{2}}}_{\rm star}\sim 1.5 \ \rm{Gyr}$. 

In the top panel of Fig. \ref{SFR_BH_acctime_uneq} we plot the corresponding
maximum black hole accretion and star formation rates. 
By defining the variable $q$ as the mass ratio
between the primary and secondary component we can fit the logarithms of the
peak BH accretion and star formation rates with the following linear relation
\begin{equation}
\log \rm{Maximum} \ [M_{\odot} \rm{yr^{-1}}]=a_{0}+a_{1} \times q,
\label{eq:uneq_fit}
\end{equation}
where $a_{0}$ and $a_{1}$ are the inferred normalization and slope respectively. 
Both the maximum BH accretion rates and star formation rates are well fitted
by Eq. \ref{eq:uneq_fit} (dotted lines in Fig. \ref{SFR_BH_acctime_uneq}) 
resulting in the following best fitting parameters respectively: 
$(a_{0,BHAR}=0.31, a_{1,BHAR}=-0.47)$ 
$(a_{0,SFR}=1.82, a_{1,SFR}=-0.24)$. The ratio of the peak BH accretion rate
to the peak star formation rate is thus of the order of $\dot{M}_{\rm
  BH,peak}/\dot{M}_{\rm SF,peak} \sim 10^{-2}$. On the other hand, the ratio of
the mean BH accretion rate to the mean star formation averaged over the whole
simulation is closer to $\dot{M}_{\rm BH,mean}/\dot{M}_{\rm SF,mean} \sim
10^{-3}$, with this ratio varying systematically between  
$\dot{M}_{\rm BH,mean}/\dot{M}_{\rm SF,mean}=2\cdot10^{-3}$ (1:1 merger) and 
$\dot{M}_{\rm BH,mean}/\dot{M}_{\rm SF,mean}=0.2\cdot10^{-3}$ (6:1
mer-ger). Interestingly, recent observations of $z\sim2$ galaxies by Daddi et
al. 2007 also find indications of an universal ratio of 
$\dot{M}_{\rm BH}/\dot{M}_{\rm SF} \sim
10^{-3}$ and this is also expected from the observed $M_{{\rm BH}}-M_{*}$
relation.

\begin{figure}
\centering 
\includegraphics[width=8.2cm]{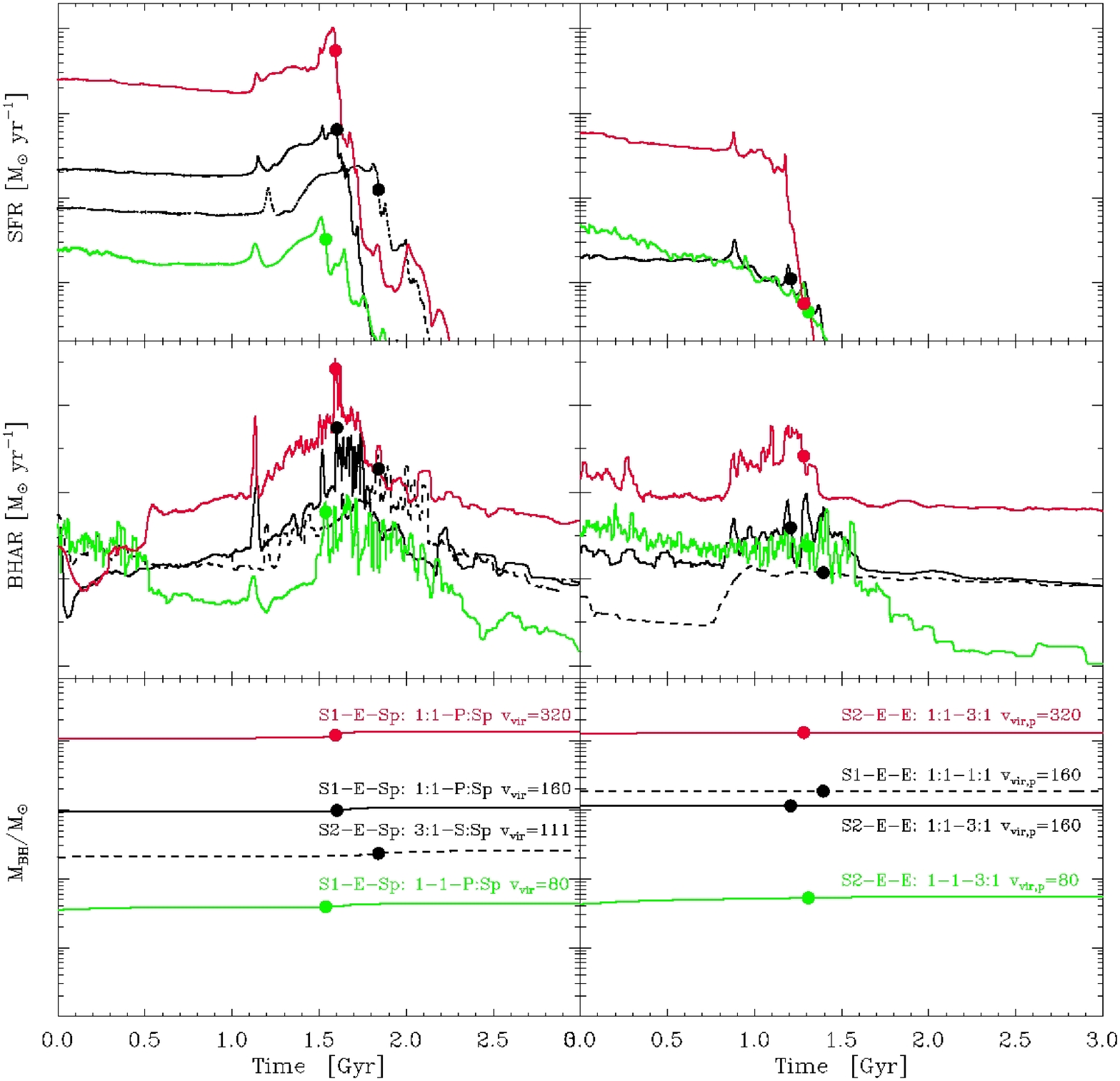}
\caption{The total star formation rate (top), the total black hole accretion rate (middle)
and the evolution of the total black hole mass (bottom) as a function of time for
four E-Sp mixed mergers \textit{(left panel)} and four E-E remergers \textit{(right panel)}.
The gas disks have initially $f_{\rm gas}=20\%$ and the virial velocities as indicated in the Figure. 
The filled circles indicate the time of merging of the BHs.}
\label{mix_EE_sfr}
\end{figure}

Finally, we analyse in Fig. \ref{mix_EE_sfr} the star formation and BH accretion rates 
for a sample of four mixed E-Sp mergers 
(left panel) and four dry E-E mergers (right panel). 
The star formation rate of the mixed E-Sp mergers
is lower than in Sp-Sp (Fig. \ref{31_sfr}) mergers, due to the lower
amount of cold gas available for star formation. The disk progenitor contains 
$f_{\rm gas}=20\%$ of gas initially, whereas the early-type progenitors 
typically have an initial gas mass fraction of $f_{\rm gas}\sim5\%$, with 
typically only $\sim1\%$ of this gas being cold and dense.
After the merger of the E-Sp galaxies the
star formation rate declines rapidly in all the merger remnants.
We define a BH mass growth factor as 
$f_{\rm BH,insitu}=\Delta M_{\rm BH,insitu}/M_{\rm BH,final}$, which gives the ratio of
BH mass growth due to gas accretion during the simulation with respect to the final BH mass.
Quantitatively, the fraction of the BH mass that accumulates by gas accretion during the mixed
mergers is in the range of $f_{\rm BH,insitu}\sim20-50\%$, with typical mean values 
of $f_{\rm BH,insitu}\sim30\%$. For the E-E mergers the initial star formation rates are generally 
very low due to the low gas fractions of $f_{\rm gas}\sim 1-5 \%$. The initial
gas fraction directly depends on the strength of the initial interaction that gave rise to
the merger remnants used as progenitors for the E-E remergers. Increasing the
masses of the progenitors and using more direct planar orbits produces
more violent initial encounters, thus decreasing $f_{\rm gas}$.
The star formation is very effectively terminated 
shortly after the merging of the E-E progenitors on comparable timescales to
1:1 Sp-Sp mergers and thus more efficiently than in 3:1 Sp-Sp and mixed E-Sp mergers.

\section{Conclusions}
\label{Conc}
In this paper we have studied the termination of star formation in merger
simulations including BH feedback. We find that the termination of star
formation by BH feedback is significantly less important for unequal-mass
disk mergers compared to equal-mass disk mergers.
The timescale for star formation termination systematically increases with
increasing progenitor mass ratios. Similarly, a systematic increase is seen in
the half-mass growth timescales of the BHs, with this timescale varying from
$\sim0.1$ Gyr for equal-mass mergers to $\sim 1$ Gyr for 6:1 mergers.
This systematic trend can be used 
as input in modeling BH accretion more realistically in semi-analytic galaxy
formation models (e.g. Croton et al. 2006).
For mass-ratios of 3:1 and higher mergers with BH feedback are unable to
completely quench the star formation, with the merger remnants showing star
formation rates roughly on the pre-merger level even 1 Gyr after completion of the merger. 
In addition, the star formation is efficiently
terminated in mixed E-Sp and dry E-E mergers 
due to the presence of the fully grown super-massive BHs 
in the early-type progenitors.

\acknowledgements
The numerical simulations were performed on the local SGI-Altix
3700 Bx2, which was partly funded by the Cluster of Excellence: ''Origin and
Structure of the Universe''.

\end{document}